\documentclass[lettersize,journal]{IEEEtran}
\usepackage{url} 
\usepackage{amsmath,amsfonts}
\usepackage{upgreek}
\usepackage{algorithmic}
\usepackage{algorithm}
\usepackage{array}
\usepackage[caption=false,font=normalsize,labelfont=sf,textfont=sf]{subfig}
\usepackage{textcomp}
\usepackage{stfloats}
\usepackage{bm}
\usepackage{verbatim}
\usepackage{graphicx}
\usepackage{cite}
\usepackage{xcolor}
\usepackage{mathtools}  
\DeclareMathOperator{\Tr}{Tr}
\usepackage{mleftright} 
\begin{document}

\title{Characterization of Nuclear Magnetism at Ultralow and Zero Field using SQUIDs}

\author{John Z. Myers, Kai Buckenmaier, Andrey N. Pravdivtsev, Markus Plaumann, and Rainer K\"orber
\thanks{J.Z. Myers is with the \textit{Physikalisch-Technische Bundesanstalt}, 10587 Berlin, Germany (email: john.myers@ptb.de). K. Buckenmaier is with the Max Planck Institute for Biological Cybernetics, 72076 T\"ubingen, Germany. A.N. Pravdivtsev is with the Molecular Imaging North Competence Center, 24118 Kiel, Germany. M. Plaumann is with the Otto-von-Guericke University Magdeburg, 39120 Magdeburg, Germany. R. K\"orber is with the \textit{Physikalisch-Technische Bundesanstalt}, 10587 Berlin, Germany.}
\thanks{Manuscript received September 25, 2024; revised xxx xx, 2024.}}

\markboth{IEEE TRANSACTIONS ON APPLIED SUPERCONDUCTIVITY}%
{Shell \MakeLowercase{\textit{et al.}}: Characterization of Nuclear Magnetism at ULF using SQUIDs}

\maketitle

\begin{abstract}

Nuclear magnetism underpins areas such as medicine in magnetic resonance imaging (MRI). Hyperpolarization of nuclei enhances the quantity and quality of information that can be determined from these techniques by increasing their signal to noise ratios by orders of magnitude. However, some of these hyperpolarization techniques rely on the use of low to ultralow magnetic fields (ULF) (nTs-mTs). The broadband character and ultrasensitive field sensitivity of superconducting quantum interference devices (SQUID) allow for probing nuclear magnetism at these fields, where other magnetometers, such as Faraday coils and flux gates do not. To this end, we designed a reactor to hyperpolarize \mbox{[1-\textsuperscript{13}C]p}yruvate with the technique, signal amplification by reversible exchange in shield enables alignment transfer to heteronuclei (SABRE-SHEATH). Hyperpolarized pyruvate has been shown to be very powerful for the diagnosis of tumours with MRI as its metabolism is associated with various pathologies. We were able to characterize the field sensitivity of our setup by simulating the filled reactor in relation to its placement in our ultralow noise, ULF MRI setup. Using the simulations, we determined that our hyperpolarization setup results in a \textsuperscript{13}C polarization of 0.4\%, a signal enhancement of $\sim$100~000~000 versus the predicted thermal equilibrium signal at earth field ($\sim$50~$\mu$T). This results in a \textsuperscript{13}C signal of 6.20$\pm$0.34~pT, which with our ultralow noise setup, opens the possibility for direct observation of the hyperpolarization and the subsequent spin-lattice relaxation without system perturbation.
\end{abstract}

\begin{IEEEkeywords}
nuclear magnetism, nuclear magnetic resonance, NMR, ultralow field, zero field, SQUIDs, hyperpolarization, parahydrogen, SABRE, SABRE-SHEATH
\end{IEEEkeywords}

\section{Introduction}
\IEEEPARstart{N}{uclear} magnetism forms the basis behind many powerful techniques, such as magnetic resonance imaging (MRI)~\cite{Hashemi2010}. In the case of MRI, when nuclei of spin $I = 1/2$ are exposed to a magnetic holding field, $B_0$, their two energy states ($\alpha$ and $\beta$) become nondegenerate. The difference in the number of nuclei occupying each state is described by their polarization (\ref{eq:1})~\cite{Atkins2023}:

\begin{equation}
\label{eq:1}
    P = \frac{N_\alpha-N_\beta}{N_\alpha+N_\beta} = \tanh\left (\frac{\gamma\hbar B_0}{2k_\mathrm{B}T}\right )
\end{equation}

\noindent where $\gamma$ is the gyromagnetic ratio of the nuclei, $\hbar$ the reduced Planck constant, $T$ the temperature, and $k_\mathrm{B}$ the Boltzmann constant.

It is this difference in occupancy of the two states that generates the nuclear magnetic moment necessary for the technique, where the signal strength is directly proportional to $P$~\cite{Green2012}. At clinically relevant $B_0$ (Ex. $B_0$~=~1.5~T, polarization is on the order of ppms, $P$~=~2.70~ppm)~\cite{Edelman2014}. While it is enough for clinical imaging, it is far from the maximum achievable signal that could be attained from higher polarization.

\begin{figure}[tbhp]
    \centering
    \includegraphics[width=\linewidth]{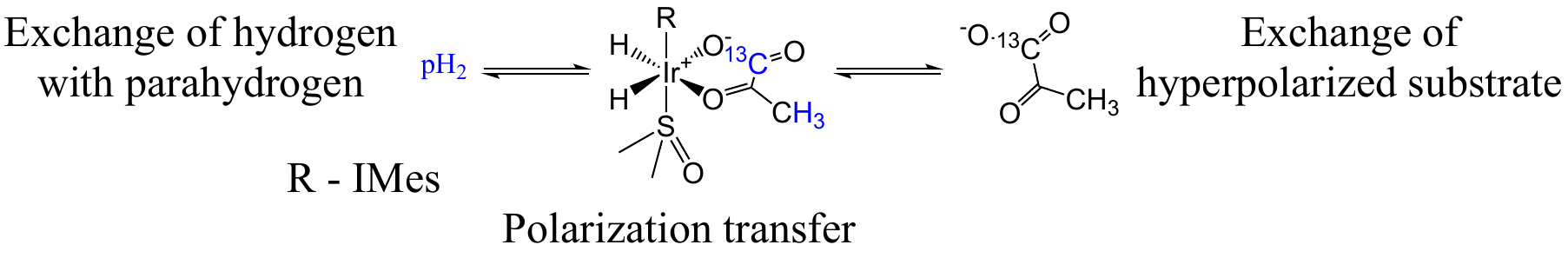}
    \caption{Schematic of hyperpolarization of \mbox{[1-\textsuperscript{13}C]p}yruvate via SABRE-SHEATH. In blue shows the transfer of parahydrogen spin order to the \textsuperscript{13}C and \textsuperscript{1}H of \mbox{[1-\textsuperscript{13}C]p}yruvate.}
    \label{fig:1}
\end{figure}

Through hyperpolarization techniques, this thermal equilibrium polarization can be increased, orders of magnitude, to the order of some percent~\cite{Green2012,ArdenkjaerLarsen2003}. One technique of interest is signal amplification by reversible exchange in shield enables alignment transfer to heteronuclei (SABRE-SHEATH), as shown in Figure~\ref{fig:1}.

Parahydrogen (pH\textsubscript{2}) is the singlet spin isomer of hydrogen, which exists as a pure state~\cite{Green2012}. Via catalysis by an iridium based catalyst, the order of this pure state can be transferred to nuclei on another molecule, such as the preclinical probe, \mbox{[1-\textsuperscript{13}C]p}yruvate~\cite{Iali2019,Cunningham2016}. This manifests as nuclear polarization, enabling metabolic imaging studies to prostate cancer diagnosis through a higher signal-to-noise ratio (SNR)~\cite{Albers2008}. However, key to this crucial polarization transfer process is a guiding field at the energy level anticrossing of the system given by (\ref{eq:2})~\cite{Pravdivtsev2013,Theis2015,Ivanov2014}:

\begin{equation}
    \label{eq:2}
    B_\mathrm{LAC} = \frac{2\pi J^{\mathrm{HH}}}{\gamma^{^1\mathrm{H}}-\gamma^{^{13}\mathrm{C}}} 
\end{equation}

\noindent where $J^{\mathrm{HH}}$ is the $J$ coupling constant between the equatorially bound hydrides on the iridium catalyst, originating from pH\textsubscript{2} (Figure \ref{fig:1}). Given a $J^{\mathrm{HH}}$~=~-10.5~Hz~\cite{Assaf2024}, this corresponds to a $B_\mathrm{LAC}$ of $\sim$330~nT, a field strength a full seven orders of magnitude weaker than that of a clinical MRI scanner. Therefore, optimization of the SABRE-SHEATH hyperpolarization process requires an understanding of nuclear spin physics at ultralow field (ULF). 

Nuclear magnetic resonance (NMR) experiments are performed by perturbing the system either through a radio frequency pulse or field switching of the $B_0$ field~\cite{Atkins2023,Barskiy2024}. The nuclear magnetic moment then precesses at a frequency proportional to its $\gamma$ given by: $\nu = \gamma B_0/2\pi$, generating an oscillating magnetic signal. This signal is typically detected with a tuned Faraday coil, with its sensitivity being directly proportional to signal frequency~\cite{Levitt2008}. This leads to intrinsically low system sensitivity at ULF~\cite{Myers2007}. This can be circumvented through use of a superconducting quantum interference device (SQUID). In contrast to a Faraday coil, a direct current SQUID operated in the flux locked loop mode has a field sensitivity independent of signal frequency, allowing for sensitive detection of low frequency to even nonoscillatory magnetic fields~\cite{Myers2007}.

Optically pumped magnetometers (OPMs) were already used in the pioneering work of Cohen Tannoudji et al.~\cite{CohenTannoudji1969} who observed the static magnetic field of hyperpolarized \textsuperscript{3}He gas. More recent OPM studies focused on detecting hyperpolarized pyruvate at ultra-low to zero field~\cite{Ortmeier2024},  investigations of its metabolic reactions~\cite{Eills2023a} and the tracking of the spin dynamics of the polarization build-up during adiabatic field sweeps across the LAC condition~\cite{Eills2024a} or by repeatedly inverting the polarization~\cite{Mouloudakis2023}.

Based on prior work from Myers et al.~\cite{Myers2024}, in this contribution we expand upon the SQUID device sensitivity and drift, demonstrating how optimizing signal enhancement obtained via SABRE-SHEATH in conjunction with an ultralow noise, ultrasensitive SQUID setup, enables the characterization of \mbox{[1-\textsuperscript{13}C]p}yruvate from ultralow down to zero field. We then show how our setup potentially enables the direct detection of nuclear magnetization without perturbation.

\begin{figure}[t]
    \centering
    \includegraphics[width=0.8\linewidth]{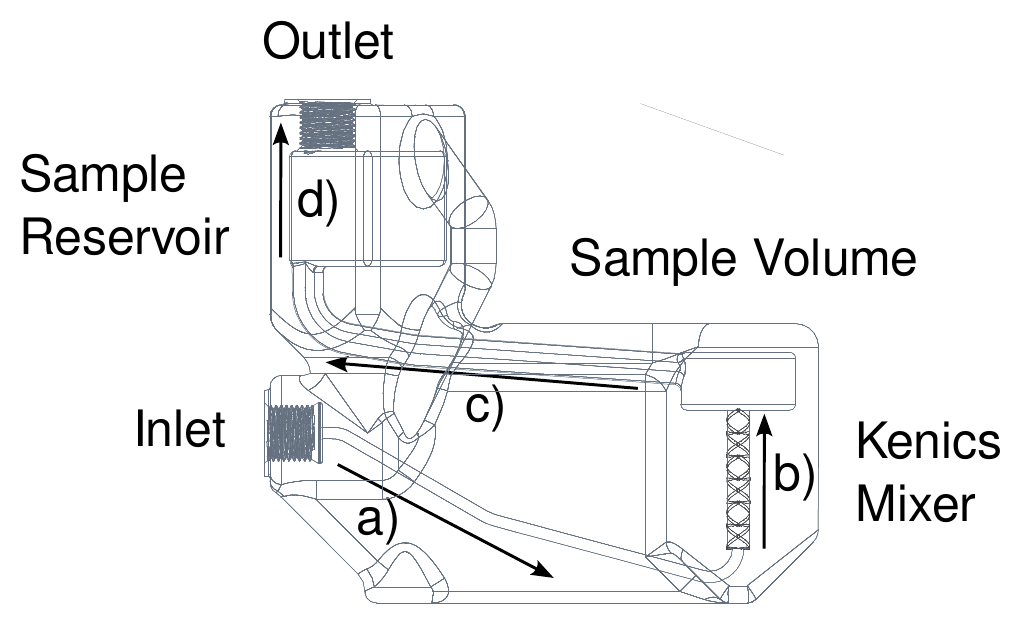}
    \caption{Schematic of SABRE reactor. a) The capillary from the inlet to the Kenics mixer. b) The Kenics static mixer (24~mm height, 3~pairs of alternating right and left-handed helices). c) Overflow channel to the sample reservoir. d) Sample reservoir and pH\textsubscript{2} outlet.}
    \label{fig:2}
\end{figure}

\section{Methods}
\subsection{SABRE-Reactor Design}
A crucial step for the optimization of signal enhancement via SABRE-SHEATH is the recipient where hyperpolarization occurs. To this end, we designed a SABRE reactor, 3D-printed using a  stereo lithography apparatus (SLA), as shown in Figure~\ref{fig:2}. pH\textsubscript{2} enters the reactor through the inlet, which is located near the midpoint of the reactor's height. This is to minimize backflow of the liquid \mbox{[1-\textsuperscript{13}C]p}yruvate solution when pH\textsubscript{2} gas is not flowing. After traveling through the inlet capillary (Figure~\ref{fig:2}a), the pH\textsubscript{2} enters the static Kenics mixer (Figure~\ref{fig:2}b)~\cite{Hobbs1997}. The Kenics mixer covers a distance of 24~mm and is composed of three pairs of alternating single turn right-handed and left-handed helices. These serve to generate turbulence between the bubbled pH\textsubscript{2} gas and liquid  sample, improving dissolution of the pH\textsubscript{2}.

The gas then enters the principal sample volume. It is a radius 10~mm~$\times$~10~mm height cylinder, which is located directly underneath the pickup coil of the SQUID setup~\cite{Myers2024}. Afterward, the hydrogen gas leaves via an overflow channel (Figure~\ref{fig:2}c) to the sample reservoir. During bubbling, \mbox{[1-\textsuperscript{13}C]p}yruvate sample also flows through the channel into the sample reservoir, due to displacement by the pH\textsubscript{2} gas. The channel is inclined to allow gravity to assist with backflow of \mbox{[1-\textsuperscript{13}C]p}yruvate sample back into the sample volume when pH\textsubscript{2} is not being bubbled.

Hydrogen gas leaves the reactor setup via the outlet located vertically above the sample reservoir (Figure~\ref{fig:2}d). This serves to allow the buoyant hydrogen gas to leave, while keeping the denser liquid \mbox{[1-\textsuperscript{13}C]p}yruvate sample inside the reactor, allowing for continuous hyperpolarization of the sample.

\subsection{Sample Preparation}
A 50~mM~\mbox{[1-\textsuperscript{13}C]p}yruvate sample was prepared with 5~mM~iridium 1,5-cyclooctadiene 1,3-bis(2,4,6-trimethyl-phenyl)imidazol-2-ylidene chloride (Ir-IMes) catalyst, and 18~mM~dimethyl sulfoxide coligand in 8~mL methanol, as described in Ref.~\cite{Myers2024}. Nominally 99\% pH\textsubscript{2} was generated in a home-built pH\textsubscript{2} generator at 25~K, based on the designs from Ref.~\cite{Buckenmaier2017}. Prior to experiments, the Ir-IMes catalyst was activated by bubbling the pyruvate solution with pH\textsubscript{2} for half an hour at 1~standard liter per hour (sLph). During experiments, pH\textsubscript{2} was bubbled at 2~sLph. 

\subsection{The SQUID system}
\begin{figure}[t]
    \centering
    \includegraphics[width=0.75\linewidth]{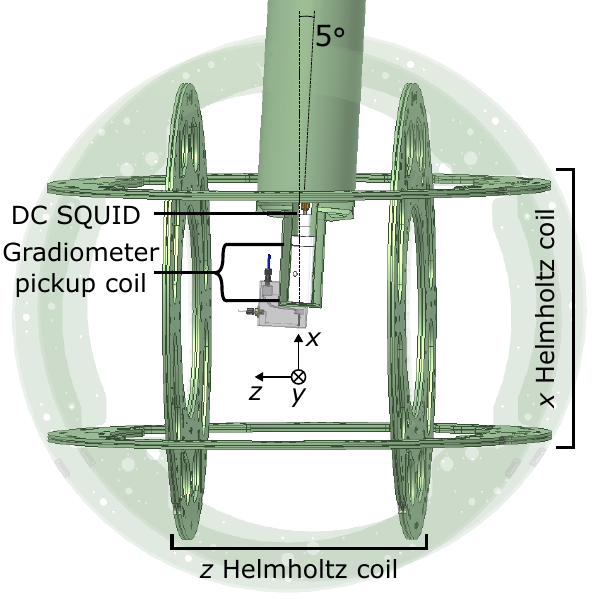}
    \caption{Schematic of SQUID-based ULF MRI setup at the PTB. Adapted from~\cite{Myers2024}.}
    \label{fig:3}
\end{figure}

ULF~NMR experiments were performed with the ULF~MRI setup within the moderately, magnetically shielded room, ZUSE-MSR, at the \textit{Physikalisch-Technische Bundesanstalt} in Berlin. A schematic of the setup is shown in Figure~\ref{fig:3}. The SABRE reactor was placed on the bottom of the ultralow noise, single channel SQUID system, operated in the LINOD2 liquid helium dewar~\cite{Storm2017}, where the dewar with the reactor were tilted by 5\textdegree ~to aid back flow of the \mbox{[1-\textsuperscript{13}C]p}yruvate solution.

The SQUID detector was composed of a single order gradiometer pickup coil (45~mm diameter, 120~mm baseline) inductively coupled into a current-sensing SQUID, sensitive along the \textit{x}-axis. Magnetic fields along the \textit{x} and \textit{z}-axes were generated with $\sim$1~m diameter Helmholtz coils.

The field sensitivity of the system was characterized by performing a 60~s acquisition at 20~kHz with the entire system in place, with no currents flowing; the acquisition was performed, during the nightly pause in train operation to minimize environmental contributions.

\subsection{ULF NMR Data Acquisition and Analysis}
\begin{figure}
    \centering
    \includegraphics[width=\linewidth]{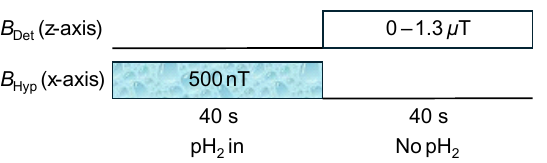}
    \caption{Schematic of ULF NMR procedure. Acquisition was performed when pH\textsubscript{2} was shunted.}
    \label{fig:4}
\end{figure}

For NMR, the \mbox{[1-\textsuperscript{13}C]p}yruvate solution was hyperpolarized via SABRE-SHEATH along the \textit{x}-axis in a 500~nT hyperpolarization field ($B_\mathrm{Hyp}$) for 40~s. This field, higher than the $B_{\mathrm{LAC}}$ was used, due to rapid chemical exchange increasing the field optimum~\cite{Adams2009}. The system was then perturbed via a nonadiabatic field switch, ramping down the $B_\mathrm{Hyp}$ field and applying a perpendicular field ($B_\mathrm{Det}$) between 0$-$1.3~$\mu$T along the \textit{z}-axis. Acquisition was then performed at sample rate of 20~kHz for 40~s with pH\textsubscript{2} being shunted away from flowing into the SABRE reactor via a three way valve (see Figure \ref{fig:4}).

During post processing, data was baseline corrected in the time domain. To eliminate dephasing, both zeroth and first order phase correction were performed on the frequency domain of the data. The initial amplitudes of the signals attributable to the \mbox{[1-\textsuperscript{13}C]p}yruvate \textsuperscript{13}C were determined by applying a Lorentzian fit to the frequency domain of the data with only zeroth order phasing for each signal (more details in Ref.~\cite{Myers2024}).

\subsection{Simulation of ULF NMR Spectra}
The ULF NMR data was compared to theory by solving the liquid state Hamiltonian of \mbox{[1-\textsuperscript{13}C]p}yruvate, as given in (\ref{eq:3})~\cite{Myers2024}:

\begin{equation}
    \label{eq:3}
    \hat{H}^{A_3X} = -\nu^A\sum_{k = 1}^3\hat{I}_z^{A_k} - \nu^X\hat{I}_z^X +J\left (\sum_{k = 1}^3\hat{\mathbf{I}}^{A_k}\cdot\hat{\mathbf{I}}^X\right )
\end{equation}

\noindent where $A$ refers to \textsuperscript{1}H, $X$ to \textsuperscript{13}C, and where $\hat{I}$ are the various product operators for the spin system and $J$ the $J$ coupling constant between the \mbox{[1-\textsuperscript{13}C]p}yruvate \textsuperscript{13}C and its \textsuperscript{1}H atoms. 

After solving the Hamiltonian, relaxation was applied to the solution phenomenologically by multiplication with a decaying first order exponential function~\cite{Myers2024}.

\subsection{Determination of Sample Polarization}
The polarization after hyperpolarization of \mbox{[1-\textsuperscript{13}C]p}yruvate by SABRE-SHEATH was determined by simulating the contribution of each point within the SABRE reactor to the detected signal, using the reactor and SQUID pickup loop geometries. This is possible, since SQUIDs directly detect magnetic flux, as opposed to change in magnetic flux~\cite{Myers2007}. Using the principal of reciprocity, the time dependent flux within the pickup loop was calculated as in (\ref{eq:4})~\cite{Hoult1979,Buckenmaier2019b}:

\begin{equation}
    \label{eq:4}
    \Phi\left (t\right ) = \mathbf{B}\left (\mathbf{r}\right )\cdot \bm{\upmu}\left (\mathbf{r},t\right )
\end{equation} 

\noindent where $\mathbf{B}\left (\mathbf{r}\right )$ is the magnetic field at point $\mathbf{r}$, produced by a unit current in the pick-up loop and $\bm{\upmu}\left (\mathbf{r},t\right )$ the magnetic moment at $\mathbf{r}$ at time $t$.

Assuming homogeneous hyperpolarization of the \mbox{[1-\textsuperscript{13}C]p}yruvate sample, $\Phi\left (t\right )$ can be reformulated as flux arising from a volume element $dV$ with magnetization $\mathbf{M}\left (t\right )$, as shown in (\ref{eq:5}):

\begin{equation}
    \label{eq:5}
    \Phi\left (t\right ) = \int_{V}\mathbf{B}\left (\mathbf{r}\right )\cdot\mathbf{M}\left (t\right )dV
\end{equation}

\noindent where $\mathbf{M}$ is related to the polarization $\mathbf{P}$. This is shown in equation (6) for the general case:

\begin{equation}
    \label{eq:6}
    \mathbf{M}=\frac{\gamma\hbar}{2}\left (N_\alpha + N_\beta\right ) \frac{\Tr\mleft(\hat{\mathbf{I}}^{X}\hat{\rho}\mright)}{\hat{I}^{X}} =\frac{\gamma\hbar}{2}\left (N_\alpha + N_\beta\right )\mathbf{P}
\end{equation}

\noindent where $\hat{\rho}$ is the density operator. Here, non-adiabatic field switching from $B_{\textrm{Hyp}}$ to $B_{\textrm{Det}}$ causes $\mathbf{M}$ and $\mathbf{P}$ to remain along the sensor sensitive $x$-axis. Using the area of the pickup loop, $A$, the effective detected field is calculated as shown in (\ref{eq:7}):

\begin{equation}
    \label{eq:7}
    B\left (t\right ) = \frac{\Phi\left (t\right )}{A} = \frac{\gamma\hbar\left (N_\alpha + N_\beta\right )}{2A}\int_{V}\mathbf{B}\left (\mathbf{r}\right )\cdot\mathbf{P}\left (t\right )dV
\end{equation}

\begin{figure}[t]
    \centering
    \includegraphics[width=\linewidth]{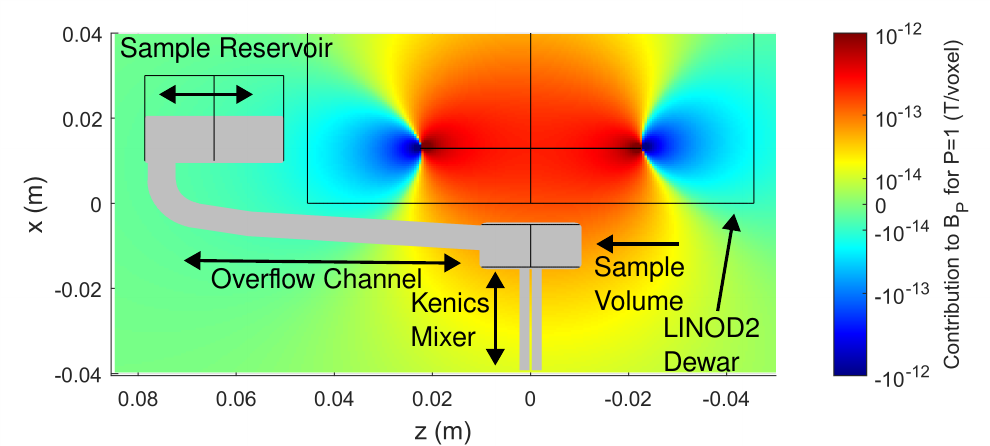}
    \caption{Field sensitivity profile of the SQUID setup for a \mbox{[1-\textsuperscript{13}C]p}yruvate sample with 100\% polarization. The filled reactor (8~mL sample with partially filled sample reservoir) is shown overlaid on the field profile.}
    \label{fig:5}
\end{figure}

\noindent $B\left (t\right )$ was then determined by discretizing the reactor volume to an isotropic 0.1~mm grid and integrating $\Phi\left (t\right )$ numerically (Figure \ref{fig:5}).

Validity of this model was confirmed by comparison with the empirical result of the \textsuperscript{1}H-signal from an 8~mL methanol sample within the reactor. This sample was thermally polarized at 5.9~mT for 50~s, ensuring thermal equilibrium. Good agreement between the observed signal of 254 fT and the simulated signal of 275 fT was found, demonstrating how knowledge of the precise geometry of the 3D-printed reactor and SQUID detection setup allows polarimetry via simulations without the need of calibration measurements.

\begin{figure}[t]
    \centering
    \includegraphics[width=\linewidth]{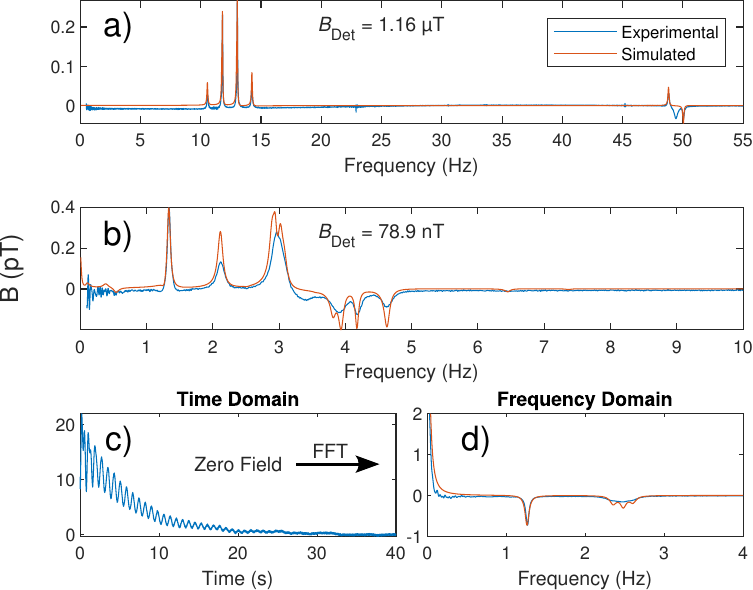}
    \caption{Selected ULF NMR results from \mbox{[1-\textsuperscript{13}C]p}yruvate hyperpolarized via SABRE-SHEATH from~\cite{Myers2024}. Simulation parameters: $J$ coupling constant:~1.223~Hz, ratio of \textsuperscript{13}C polarization to \textsuperscript{1}H:~5~000, relaxation constant:~0.02~s\textsuperscript{-1}. a) Spectrum from $B_{\mathrm{Det}}$~=~1.16~$\mu$T. The emissive peak at $\sim$50~Hz stems from hyperpolarized othohydrogen~\cite{Barskiy2014, Buckenmaier2019} which was not considered in the simulation. b) Spectrum from $B_{\mathrm{Det}}$~=~78.9~nT. c) Baseline corrected time domain of zero field data. d) Spectrum at zero field.}
    \label{fig:6}
\end{figure}

\section{Results}
\subsection{Zero to ULF NMR}
Selected NMR spectra from the application of our ultralow noise ULF MRI SQUID setup to the system of \mbox{[1-\textsuperscript{13}C]p}yruvate hyperpolarized by SABRE-SHEATH are shown in Figure \ref{fig:6}. Using our setup and hyperpolarization, it was possible to characterize the NMR spectra of \mbox{[1-\textsuperscript{13}C]p}yruvate from ultralow to zero field and compare the empirical results with what was predicted from simulations. Not only did the simulations give results that agreed with the experimental data at the weak (Figure \ref{fig:6}a) and strong coupling regimes (Figure \ref{fig:6}d), but they also agreed with the features observed in the experimental data in the intermediate coupling regime (Figure \ref{fig:6}b)~\cite{Myers2024}.

\subsection{\mbox{[1-\textsuperscript{13}C]P}yruvate Sample Polarization}
Upon summing the voxels corresponding to the SABRE reactor volume in Figure \ref{fig:5}, for a 100\% polarized \mbox{[1-\textsuperscript{13}C]p}yruvate sample, the expected signal was determined to be 1.90~nT. One feature of note upon summation of the voxels is that the sample in the sample reservoir has a negative contribution to the total signal achieved. In practice, this results in a $\sim$10\% decrease in the total magnetic flux from the sample in the pickup coil.

After summation of the initial amplitudes of the signals attributable to \textsuperscript{13}C from the hyperpolarized \mbox{[1-\textsuperscript{13}C]p}yruvate, the total \textsuperscript{13}C signal was determined to be 6.20$\pm$0.34~pT, yielding a total \textsuperscript{13}C polarization of $\sim$0.4\%.

\subsection{ULF MRI SQUID Setup Field Drift}
\begin{figure}[htbp]
    \centering
    \includegraphics[width=\linewidth]{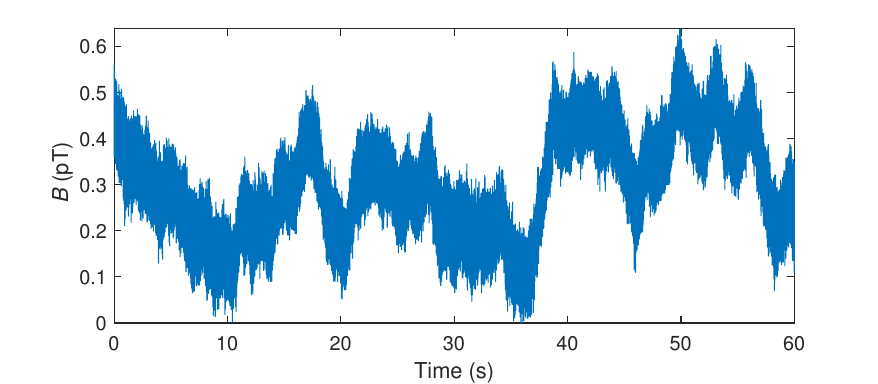}
    \caption{Field drift of the ULF MRI SQUID setup.}
    \label{fig:7}
\end{figure}

A field drift of $\sim$0.6~pT was recorded by the SQUID setup over a one minute period as shown in Figure \ref{fig:7}. In comparison to the 6.20$\pm$0.34~pT attained by SABRE-SHEATH hyperpolarization of \mbox{[1-\textsuperscript{13}C]p}yruvate, the signal was $\sim$10 times higher than the drift.

\section{Discussion and Conclusions}
The application of our ultralow noise ULF MRI SQUID setup allowed for the characterization of the evolution of NMR signals from \mbox{[1-\textsuperscript{13}C]p}yruvate hyperpolarized via SABRE-SHEATH from ULF to zero field. This was made possible through the design and implementation of our home-built SABRE reactor, which used features from a static Kenics mixer to an overflow reservoir. This enabled the ability to optimize sample mixing and stabilized the sample height to maximize the attainable signal achieved from SABRE-SHEATH hyperpolarization. From this setup, we were able to attain a \textsuperscript{13}C signal of 6.20$\pm$0.34~pT, which upon simulation of the field contributions of the SABRE reactor, resulted in a total polarization of $\sim$0.4\%. The sensitivity of our setup should allow a direct measurement of polarization build-up without any field switching or radio frequency pulses and might reveal polarization oscillations during coherent spin order transfer of the SABRE process as observed during adiabatic field ramps~\cite{Eills2024a}. In addition, ULF MRI of \textsuperscript{13}C in \mbox{[1-\textsuperscript{13}C]p}yruvate, as demonstrated by Kempf et al.~\cite{Kempf2024} could be used for future studies and optimizations of gas-fluid dynamics in the reaction chamber without susceptibility distortions.

\section*{Acknowledgments}
We would like to thank Michael Tayler and James Eills who brought to our attention previous work by Cohen-Tannoudji.

This work was funded by the DFG under Grant No. 469366436. ANP acknowledges support from the German Federal Ministry of Education and Research (BMBF) within the framework of the e:Med research and funding concept (01ZX1915C), DFG (555951950, 527469039). MOIN CC was founded by a grant from the European Regional Development Fund (ERDF) and the Zukunftsprogramm Wirtschaft of Schleswig-Holstein (Project no. 122-09-053).

\bibliographystyle{IEEEtran}
\bibliography{main_R1a}

\newpage
\vfill

\end{document}